\documentclass[aps,%
12pt,%
notitlepage,%
final,%
oneside,%
onecolumn,%
nobibnotes,%
nofootinbib,%
superscriptaddress,%
noshowpacs,%
centertags,%
amsmath,amssymb,%
a4paper]{revtex4}

\begin{document}

\title{The Rarita--Schwinger field: renormalization and phenomenology}
\author{\firstname{A.E.} \surname{Kaloshin}}
\email{kaloshin@physdep.isu.ru}
\author{\firstname{V.P.} \surname{Lomov}}
\email{lomov@physdep.isu.ru}

\affiliation{Irkutsk State University, Russia}

\begin{abstract}
  We discuss renormalization of propagator of interacting Rarita--Schwinger field. Spin-$3/2$
  contribution after renormalization takes usual resonance form. For non-leading spin-$1/2$
  terms we found procedure, which guarantees absence of poles in energy plane. The obtained
  renormalized propagator has one free parameter and is a straight generalization of the famous
  free propagator of Moldauer and Case. Application of this propagator for production of
  $\Delta^{++}(1232)$ in $\pi^{+}\particle{p}\to \pi^{+}\particle{p}$ leads to good description
  of total cross-section and to reasonable agreement with results of partial wave analysis.
\end{abstract}

\maketitle


\section{Introduction}

The vector-spinor Rarita--Schwinger (R.--S.) field is used to describe spin $3/2$ particles in
quantum field theory \cite{RS1941}.  It is known that all theories with higher spins $s
\geqslant 1$ are faced with a common problem: the corresponding field contains extra degrees of
freedom related with non-leading spins $s-1, \dots$. Therefore some constrains should be
imposed on the free field to exclude these degrees of freedom.  But inclusion of interaction
usually breaks these constrains and it generates the main problems. These issues for
spin-$3/2$ were discussed in \cite{JS1961}, \cite{VZ1969} and in numerous following
papers. Problem of consistence of constrains and interactions is discussed actively up to now
\cite{KT1987,Pas1999,KA2001,Pas2000}.

The main applications of Rarita--Schwinger formalism are related with baryon spectroscopy. But
because of old theoretical problems any application for spin-$3/2$ baryons with necessity contains
some approximations.

We prefer to investigate the dressed propagator of \RSf instead of equations of motion and
constrains. Besides technical advantages, it allows to apply the obtained dressed propagator to
description of experimental data. The full non-renormalized propagator of \RSf with account of
all spin components was found in \cite{KL2004,KL2006}. In this paper we study the procedure of
renormalization of this propagator. After it we apply the constructed renormalized propagator to
production of $\Delta^{++}(1232)$ in $\pi^{+}\particle{p}\to \pi^{+}\particle{p}$ reaction.



\section{Dressed propagator of Raria--Schwinger field}

\subsection{Most general lagrangian of \RSf}

Free lagrangian of the \RSf is defined by differential operator $S^{\mu\nu}$ which is, in fact,
the inverse propagator
\begin{equation} \label{eq:lagr}
  \Lagr=\overline{\Psi}\vphantom{\Psi}^{\mu}S_{\mu\nu}\Psi^{\nu}.
\end{equation}
The standard form of $S^{\mu\nu}$ is (see e.g. \cite{MC1956})
\begin{equation} \label{eq:stand}
  S^{\mu\nu} = (\hat{p}-M)g^{\mu\nu}+A(\gamma^{\mu}p^{\nu}+\gamma^{\nu}p^{\mu})+\frac{1}{2}(3A^2+2A+1)\gamma^{\mu}
             \hat{p}\gamma^{\nu}+M(3A^2+3A+1)\gamma^{\mu}\gamma^{\nu}.
\end{equation}
Here $M$ is the mass of spin-$3/2$ particle, $p_{\mu}=\imath\partial_{\mu}$ and $A$ is an
arbitrary real parameter. Equations of motion, following from \eqref{eq:stand}, lead to
constrains $p\Psi=\gamma\Psi=0$ and therefore to exclusion of the spin-$1/2$ degrees of
freedom.  In other words, the corresponding terms in propagator should not have poles in energy
plane.

We want to derive the most general form of the \RS lagrangian without additional assumption
about the nature of spin-$1/2$ contributions. Depending on the choice of parameters the $s=1/2$
components may have or have not the poles in the complex energy plane. Such construction will be
useful at renormalization of the dressed \RSf propagator even if we require the spin-$1/2$
components to be unphysical.

Let us formulate the main requirements for lagrangian:
\begin{enumerate}
\item The fermion lagrangian is linear with respect to derivatives.
\item It should be hermitian $\Lagr^{\dag}=\Lagr$ or $\gamma^{0} (S^{\mu\nu})^{\dag}\gamma^{0}=S^{\nu\mu}$.
\item The spin-$3/2$ contribution has standard pole form (to be specified below).
\item Lagrangian should not be singular at $p^{2}\to 0$. This point is rather evident but it happens
  that some rough methods generate singularities in a propagator (see e.g. discussion in
  \cite{BDM1989}).
\end{enumerate}

The suitable starting point to construct the generalized lagrangian is decomposition of
$S^{\mu\nu}$ in $\gamma$-matrix basis%
\footnote{%
  We use conventions of Bjorken and Drell textbook \cite{Bjo}: $\epsilon_{0123}=1$,
  $\gamma^{5}=\gamma_{5}=\imath\gamma^{0}\gamma^{1}\gamma^{2}\gamma^{3}$ except that
  $\sigma^{\mu\nu}=\frac{1}{2}[\gamma^{\mu},\gamma^{\nu}]$.%
}
\eqref{eq:app-eq1} with arbitrary coefficients. The first requirement remains $6$ complex
coefficients in \eqref{eq:app-eq1}
\begin{equation} \label{eq:eq4}
  \begin{split}
    S^{\mu\nu} &= g^{\mu\nu}\cdot s_{1}+\hat{p}g^{\mu\nu}\cdot s_{4}+p^{\mu}\gamma^{\nu}\cdot
    s_{5}+\gamma^{\mu}p^{\nu}\cdot s_{6}+\sigma^{\mu\nu}\cdot s_{7}+
    \imath\epsilon^{\mu\nu\lambda\rho}\gamma_{\lambda}\gamma^{5} p_{\rho}\cdot s_{10}=\\
    &=g^{\mu\nu}(s_{1}-s_{7}) +
    \hat{p}g^{\mu\nu}(s_{4}-s_{10})+p^{\mu}\gamma^{\nu}(s_{5}+s_{10})+
    \gamma^{\mu}p^{\nu}(s_{6}+s_{10})+\gamma^{\mu}
    \gamma^{\nu}s_{7}-\gamma^{\mu}\hat{p}\gamma^{\nu}s_{10}.
  \end{split}
\end{equation}
If we start from the $\gamma$-matrix decomposition with non-singular coefficients the fourth
requirement is fulfilled automatically.

This expression satisfies the condition $\gamma^{0}(S^{\mu\nu})^{\dag}\gamma^{0}=S^{\nu\mu}$, if
$s_{1}$, $s_{4}$, $s_{7}$, $s_{10}$ are real parameters while $s_{5}$ and $s_{6}=s_{5}^{*}$ may
be complex. It is convenient to introduce the new notations
\begin{equation*}
  s_{1}=r_{1},\quad s_{4}=r_{4},\quad s_{7}=r_{7},\quad
  s_{10}=r_{10},\quad s_{5}=r_{5}+ \imath a_{5},\quad s_{6}=r_{5}- \imath
  a_{5},
\end{equation*}
where all parameters are real.

To take into account the third requirement, we need to recognize the spin-$3/2$ part of inverse
propagator. It is easy to do in the $\hat{p}$-basis (see Appendix~\ref{sec:app-pb} for details)
\begin{equation} \label{eq:eq5}
  S^{\mu\nu}=(\hat{p}-M)\left(\operP^{3/2}\right)^{\mu\nu}+(\text{spin-$1/2$ contributions}).
\end{equation}
Reversing the Eq.~\eqref{eq:eq5}, we obtain propagator with the standard pole behavior of
spin-$3/2$ contribution
\begin{equation} \label{eq:eq6}
  G^{\mu\nu} = \frac{1}{\hat{p}-M}\left(\operP^{3/2}\right)^{\mu\nu}+(\text{spin-$1/2$ contributions}).
\end{equation}
Eq.~\eqref{eq:eq5} gives%
\footnote{%
  We need to use different bases, so to distinguish them we use different notations: $s_{i}$,
  $S_{i}$, $\bar{S}_{i}$ for coefficients in $\gamma$-, $\hat{p}$- and $\Lambda$-basis
  respectively. Technical details are collected in Appendix~\ref{sec:app-sp-decomp}.%
}
(using formulae \eqref{eq:app-eq4}, \eqref{eq:app-eq5} for transition from one basis to another)
\begin{equation}
  S_{1}=s_{1}-s_{7}=-M, \quad S_{2}=s_{4}-s_{10}=1 .
\end{equation}
So the $r_{7}$, $r_{10}$ are dependent values
\begin{equation*}
  \begin{split}
    r_{7}=M + r_{1}, \quad r_{10}=r_{4}-1
  \end{split}
\end{equation*}
and we come to four-parameter $(r_{1},r_{4},r_{5},a_{5})$ lagrangian which satisfies all the necessary requirements.
\begin{equation} \label{eq:mostl}
  S^{\mu\nu}=g^{\mu\nu}(\hat{p}-M) + p^{\mu}\gamma^{\nu}(r_{5}+ r_{4}-1 +\imath a_{5}) +
            p^{\nu}\gamma^{\mu}(r_{5}+r_{4}-1 -\imath a_{5})+\gamma^{\mu}\gamma^{\nu}(M+r_{1})-
            \gamma^{\mu}\hat{p}\gamma^{\nu}(r_{4}-1).
\end{equation}

To build the propagator of the \RSf we need to reverse \eqref{eq:mostl}
\begin{equation}
  G^{\mu\nu}(p) = \big(S^{-1}\big)^{\mu\nu},
\end{equation}
and the $\Lambda$-basis is convenient here, see eq. \eqref{eq:solve} below.

Let us write down the denominators following from our lagrangian
\eqref{eq:mostl}:
\begin{align*}
  \Delta_{1}(W) &= -M(3M+4r_{1})+ 2W(Mr_{5}-Mr_{4}-r_{1})+W^{2}(-3a_{5}^{2}-3(r_{4}+r_{5})^{2}+4r_{5}+2r_{4}),\\
  \Delta_{2}(W) &= \Delta_{1}(W\to-W).
\end{align*}
The requirement the spin-$1/2$ contribution to be unphysical is equivalent to condition
$\Delta_{1}=\const$ and we come to equations
\begin{equation} \label{eq:nonph}
  \begin{split}
    M(r_{5}-r_{4})-r_{1} &= 0,\\
    3a_{5}^{2}+3(r_{4}+r_{5})^{2}-4r_{5}-2r_{4} &= 0.
  \end{split}
\end{equation}
One can rewrite it in terms of sum and difference $\sigma=r_{5}+r_{4}$, $\delta=r_{5}-r_{4}$
\begin{equation} \label{eq:nonph'}
  \begin{split}
    r_{1} &= M \delta,\\
    \delta&= 3(\sigma^{2}-\sigma+a_{5}^{2}) .
  \end{split}
\end{equation}

There exists the well-known transformation of \RSf
\begin{equation} \label{eq:RSf-repar}
  \Psi_{\mu} \to \Psi^{\prime}_{\mu} : \quad \Psi_{\mu}=\theta_{\mu\nu}(B)\Psi^{\prime\nu},
\end{equation}
where $\theta_{\mu\nu}(B)=g_{\mu\nu}+B\gamma_{\mu}\gamma_{\nu}$ and $B=b+\imath\beta$ is a complex
parameter.

This transformation doesn't touch the spin-$3/2$ because $(\operP^{3/2})^{\mu\nu}$ operator is
orthogonal to $\gamma^{\alpha}$. If one apply it to the inverse propagator \eqref{eq:mostl}
\begin{equation}\label{eq:trS}
  S_{\mu\nu}\to  S^{\prime}_{\mu\nu}=\theta_{\mu\alpha}(B^{*}) S^{\alpha\beta}\theta_{\beta\nu}(B) ,
\end{equation}
one can see that $S^{\prime}_{\mu\nu}$ keeps all the properties of $S_{\mu\nu}$
\eqref{eq:mostl}. It means that, in fact, we have reparametrization -- after transformation of
the \eqref{eq:mostl} we obtain the same propagator with changed parameters
\begin{equation} \label{eq:eq10}
  \theta_{\mu\alpha}(B^{*})S^{\alpha\beta}(r_{1},r_{4},r_{5},a_{5})
  \theta_{\beta\nu}(B)=S_{\mu\nu}(r_{1}^{\prime},r_{4}^{\prime},r_{5}^{\prime},a_{5}^{\prime}).
\end{equation}

Using the field renormalization $\Psi^{\prime}_{\mu}=\theta_{\mu\nu}(B)\Psi^{\nu}$, we can
eliminate two of four parameters in free lagrangian \eqref{eq:mostl}. After renormalization the
$\theta$-factor will appear in vertex.

Recall that the standard interaction lagrangian of $\pi\particle{N}\Delta$ for \RSf contains
this $\theta$-factor
\footnote{
  Below we suppose as usual parameter in vertex to be real, so $a_{5}=0$ in \eqref{eq:mostl}
}
\begin{equation}
  \label{eq:interac-lagr}
  \LagrInt=g_{\pi\particle{N}\Delta}\bar{\Psi}\vphantom{\Psi}_{\mu}(x)\theta^{\mu\nu}(a)\Psi(x)\cdot\partial_{\nu}\phi(x) + \hc.
\end{equation}
Details of transformation can be found in \cite{KLM2005}. Let us mention also the previous
works \cite{Munczek1967,FY1973} where generalized \RS lagrangian was discussed.

\subsection{Dyson--Schwinger equation}

The Dyson--Schwinger equation for the propagator of the \RSf has the following form
\begin{equation}
  G^{\mu\nu}=G^{\mu\nu}_{0}+G^{\mu\alpha}\Sigma_{\alpha\beta}G^{\beta\nu}_{0}.
\end{equation}
Here $G_{0}^{\mu\nu}$ and $G^{\mu\nu}$ are the free and full propagators respectively,
$\Sigma^{\mu\nu}$ is a self-energy contribution. The equation may be rewritten for inverse
propagators as
\begin{equation} \label{eq:inverD}
  (G^{-1})^{\mu\nu}=(G^{-1}_{0})^{\mu\nu}-\Sigma^{\mu\nu}.
\end{equation}
If we consider the self-energy $\Sigma^{\mu\nu}$ as a known value (so called "rainbow"
approximation) then the problem is reduced to reversing of relation \eqref{eq:inverD}.

The most convenient basis ($\Lambda$-basis) for the spin-tensor $S^{\mu\nu}(p)$ is build with
use of five known tensor operators \cite{Nie1981,BDM1989,PS1995}
\begin{equation} \label{eq:oper}
  \begin{split}
    (\mathcal{P}^{3/2})^{\mu\nu}&=g^{\mu\nu}-n_{1}^{\mu}n_{1}^{\nu}-n_{2}^{\mu}n_{2}^{\nu}\\
    (\mathcal{P}^{1/2}_{11})^{\mu\nu}&=n_{1}^{\mu}n_{1}^{\nu}, \qquad
    (\mathcal{P}^{1/2}_{22})^{\mu\nu}=n_{2}^{\mu}n_{2}^{\nu},  \\
    (\mathcal{P}^{1/2}_{21})^{\mu\nu}&=n_{1}^{\mu}n_{2}^{\nu}, \qquad
    (\mathcal{P}^{1/2}_{12})^{\mu\nu}=n_{2}^{\mu}n_{1}^{\nu}.
  \end{split}
\end{equation}
Here we introduced the unit "vectors"
\begin{equation} \label{eq:sv-uv}
  n_{1}^{\mu}=\frac{1}{\sqrt{3}p^{2}}(-p^{\mu}+\gamma^{\mu}\hat{p})\hat{p}, \quad
  n_{2}^{\mu}=p^{\mu}/\sqrt{p^2}I_{4}, \quad
  (n_{i}\cdot n_{j})=\delta_{ij}I_{4},
\end{equation}
$I_{4}$ is unit $4\times 4$ matrix. To build a basis with good multiplicative properties we need
also the off-shell projection operators $\Lambda^{\pm}=(1\pm\hat{p}/\sqrt{p^{2}})/2$. Ten elements
of $\Lambda$-basis look as
\begin{flalign} \label{eq:L-basis}
  \mathcal{P}_{1} &= \Lambda^{+}\mathcal{P}^{3/2},\,&
  \mathcal{P}_{3} &= \Lambda^{+}\mathcal{P}^{1/2}_{11},\,&
  \mathcal{P}_{5} &= \Lambda^{+}\mathcal{P}^{1/2}_{22},\,&
  \mathcal{P}_{7} &= \Lambda^{+}\mathcal{P}^{1/2}_{21},\,&
  \mathcal{P}_{9} &= \Lambda^{+}\mathcal{P}^{1/2}_{12},\notag \\
  \mathcal{P}_{2} &= \Lambda^{-}\mathcal{P}^{3/2},\,&
  \mathcal{P}_{4} &= \Lambda^{-}\mathcal{P}^{1/2}_{11},\,&
  \mathcal{P}_{6} &= \Lambda^{-}\mathcal{P}^{1/2}_{22},\,&
  \mathcal{P}_{8} &= \Lambda^{-}\mathcal{P}^{1/2}_{21},\,&
  \mathcal{P}_{10}&= \Lambda^{-}\mathcal{P}^{1/2}_{12},
\end{flalign}
where tensor indices are omitted.

Decomposition of a spin-tensor in this basis has the following form:
\begin{equation} \label{eq:l-expan}
  S^{\mu\nu}(p)=\sum_{i=1}^{10}\mathcal{P}^{\mu\nu}_{i}\bar{S}_{i}(p^{2}).
\end{equation}
The $\Lambda$-basis has very simple multiplicative properties which are represented in the
Table~\ref{tab:tt}.

Let us denote the inverse dressed and free propagators by $S^{\mu\nu}$ and $S^{\mu\nu}_{0}$
respectively. Decomposing the $S^{\mu\nu}$, $S_{0}^{\mu\nu}$ and $\Sigma^{\mu\nu}$ in
$\Lambda$-basis according to \eqref{eq:l-expan} we reduce the equation \eqref{eq:inverD} to set of
equations for the scalar coefficients
\begin{equation*}
  \bar{S}_{i}(p^2)=\bar{S}_{0i}(p^2)-\bar{\Sigma}_{i}(p^2), \quad i=1\dots 10 .
\end{equation*}

After that the reversing of the $S^{\mu\nu}$ leads to equations for the coefficients
$\bar{G}_{i}$:
\begin{equation} \label{eq:ddecomp}
  \Big(\sum_{i=1}^{10}\mathcal{P}^{\mu}_{i\alpha}\cdot\bar{G}^{i}(p^2)\Big)\cdot
  \Big(\sum_{k=1}^{10}\mathcal{P}^{\alpha\nu}_{k}\cdot\bar{S}^{k}(p^2)\Big)=
  \sum_{i=1}^{6}\mathcal{P}^{\mu\nu}_{i} ,
\end{equation}
which are easy to solve due to simple multiplicative properties of $\mathcal{P}^{\mu\nu}_{i}$:
\begin{flalign} \label{eq:solve}
   \bar{G}_{1} &= 1/\bar{S}_{1}, \quad
  &\bar{G}_{3} &= \bar{S}_{6}/\Delta_{1}, \quad
  &\bar{G}_{5} &= \bar{S}_{4}/\Delta_{2}, \quad
  &\bar{G}_{7} &=-\bar{S}_{7}/\Delta_{1},\quad
  &\bar{G}_{9} &= -\bar{S}_{9}/\Delta_{2}, \notag\\
   \bar{G}_{2} &= 1/\bar{S}_{2}, \quad
  &\bar{G}_{4} &= \bar{S}_{5}/\Delta_{2}, \quad
  &\bar{G}_{6} &= \bar{S}_{3}/\Delta_{1}, \quad
  &\bar{G}_{8} &=-\bar{S}_{8}/\Delta_{2},\quad
  &\bar{G}_{10}&=-\bar{S}_{10}/\Delta_{1},
\end{flalign}
where $\Delta_{1}=\bar{S}_{3}\bar{S}_{6}-\bar{S}_{7}\bar{S}_{10}$,
$\Delta_{2}=\bar{S}_{4}\bar{S}_{5}-\bar{S}_{8}\bar{S}_{9}$.

The $\bar{G}_{1}$, $\bar{G}_{2}$ terms which describe the spin-$3/2$ have the usual resonance
form, the $\bar{G}_{3}$ -- $\bar{G}_{10}$ terms correspond to the spin-$1/2$ contributions.

\subsection{Matrix form of spin-$1/2$ sector}
\label{sec:2x2-mat}

The spin-$1/2$ related terms of \RSf in $\Lambda$-basis can be written in following form
\begin{equation} \label{eq:spin12}
  \begin{split}
    G^{\mu\nu}_{s=1/2}(p)&=\sum_{i=3}^{10} \operP_{i}^{\mu\nu} \bar{G}_{i}= \\
    &=(\bar{G}_{3}\Lambda^{+} + \bar{G}_{4}\Lambda^{-}) \big(\operP^{1/2}_{11} \big)^{\mu\nu} +
    (\bar{G}_{5}\Lambda^{+} + \bar{G}_{6}\Lambda^{-}) \big( \operP^{1/2}_{22} \big)^{\mu\nu}+ \\
    &+(\bar{G}_{7}\Lambda^{+} + \bar{G}_{8}\Lambda^{-}) \big( \operP^{1/2}_{21} \big)^{\mu\nu} +
    (\bar{G}_{9}\Lambda^{+} + \bar{G}_{10}\Lambda^{-}) \big( \operP^{1/2}_{12} \big)^{\mu\nu}= \\
    &=(\bar{G}_{3}\Lambda^{+} + \bar{G}_{4}\Lambda^{-})n_{1}^{\mu}n_{1}^{\nu}+
    (\bar{G}_{5}\Lambda^{+} +
    \bar{G}_{6}\Lambda^{-})n_{2}^{\mu}n_{2}^{\nu}+ \\
    &+(\bar{G}_{7}\Lambda^{+} + \bar{G}_{8}\Lambda^{-})n_{2}^{\mu}n_{1}^{\nu}
    +(\bar{G}_{9}\Lambda^{+} + \bar{G}_{10}\Lambda^{-})n_{1}^{\mu}n_{2}^{\nu}
  \end{split}
\end{equation}

Now it is convenient to shift $\Lambda^{\pm}$ in \eqref{eq:spin12} to be between the unit
vectors%
\footnote{%
  There are useful properties: $\Lambda^{\pm}n_{2}^{\mu}=n_{2}^{\mu}\Lambda^{\pm}$,
  $\Lambda^{\pm}n_{1}^{\mu}=n_{1}^{\mu}\Lambda^{\mp}$%
}
$n_{i}^{\mu}\Lambda^{\pm} n_{j}^{\nu}$

After that we come to typical matrix form for mixing of two propagators where appeared two
matrices $2\times 2$ accompanied by $\Lambda^{\pm}$ projectors
\begin{equation} \label{eq:matr}
  \begin{split}
    G^{\mu\nu}_{s=1/2}(p) &= \left(
      \begin{array}{cc}
        n_{1}^{\mu}\Lambda^{-} & n_{2}^{\mu}\Lambda^{-}
      \end{array} \right)
    \left(
      \begin{array}{cc}
        \bar{G}_{3}  & \bar{G}_{7} \\
        \bar{G}_{10} & \bar{G}_{6}
      \end{array} \right)
    \left(
      \begin{array}{c}
        \Lambda^{-}n_{1}^{\nu}  \\
        \Lambda^{-}n_{2}^{\nu}
      \end{array} \right) + \\[2mm]
    &+ \left(
      \begin{array}{cc}
        n_{1}^{\mu}\Lambda^{+} &  n_{2}^{\mu}\Lambda^{+}
      \end{array} \right)
    \left(
      \begin{array}{cc}
        \bar{G}_{4} & \bar{G}_{8} \\
        \bar{G}_{9} & \bar{G}_{5}
      \end{array} \right)
    \left(
      \begin{array}{c}
        \Lambda^{+}n_{1}^{\nu}  \\
        \Lambda^{+}n_{2}^{\nu}
      \end{array} \right) \\
    &\equiv T^{\mu\nu} [G^{+}, G^{-}]
  \end{split}
\end{equation}
Here we introduce special notation for this object. Matrix form \eqref{eq:matr} has very
convenient property at multiplication: if we have spin-tensors $A^{\mu\nu}(p)$ (characterized by $A^{+},
A^{-}$ matrices) and $B^{\mu\nu}(p)$ (characterized by $B^{+}, B^{-}$) then multiplication of
these spin-tensors is reduced to multiplication of corresponding $2\times 2$ matrices.
\begin{equation} \label{eq:mul}
  T^{\mu\rho} [A^{+}, A^{-}]\cdot T_{\rho}\vphantom{T}^{\nu} [B^{+}, B^{-}]= T^{\mu\nu} [A^{+}B^{+}, A^{-}B^{-}].
\end{equation}
In particular, the $\theta$-transformation also can be written in
matrix form:
\begin{equation}
  \theta^{\mu\nu}(a)\equiv g^{\mu\nu}+ a \gamma^{\mu}\gamma^{\nu}=
  \theta^{\mu\nu}_{s=3/2} + T^{\mu\nu} [\theta^{+}, \theta^{-}],
\end{equation}
where
\begin{equation}
  \theta^{+} = \left(
    \begin{array}{cc}
        1+3a    & \sqrt{3}a \\
      \sqrt{3}a &   1+a
    \end{array}\right)
  ,\quad
  \theta^{-} = \left(
    \begin{array}{cc}
         1+3a    & -\sqrt{3}a \\
      -\sqrt{3}a &   1+a
    \end{array} \right).
\end{equation}



\section{Renormalization of \RS propagator}

\subsection{Renormalization of spin-$3/2$ terms}

It is convenient to renormalize the inverse propagator $S^{\mu\nu}$. Spin-$3/2$ contribution
corresponds to first two term in \eqref{eq:l-expan}
\begin{equation}
  S^{\mu\nu}_{3/2}=\sum_{i=1}^{2} \operP_{i}^{\mu\nu} \bar{S}^{i},
\end{equation}
where
\begin{equation}
  \bar{S}_{1}(W)= W-M-\bar{\Sigma}_{1}(W)= W-M-\Big(\Sigma_{1}\big(W^{2}\big)+W\Sigma_{2}\big(W^{2}\big)\Big)
\end{equation}
and $\bar{\Sigma}_{2}(W)=\bar{\Sigma}_{1}(-W)$.

If to use the on-mass-shell scheme of renormalization then $M$ is the renormalized mass. For
resonance state located higher the threshold, the renormalization condition is
\begin{equation} \label{eq:renorm}
  \bar{S}_{1} = W-M+o(W-M)+\imath\frac{\Gamma}{2} \quad\text{at}\quad W \sim M .
\end{equation}
Note that real part of \eqref{eq:renorm} is some requirement on the subtraction constants of
self-energy functions $\Sigma_{1}$, $\Sigma_{2}$, whereas the imaginary part simply relates the
coupling constant and width.

So for normalization we should subtract the self-energy contribution twice at this point
\begin{equation}
  \bar{S}_{1}^r(W)= W-M-\big[\bar{\Sigma}_{1}(W)-\Re\bar{\Sigma}_{1}(M)-\Re \bar{\Sigma}_{1}^{\prime}(M)(W-M)\big]
\end{equation}
and after it $\bar{\Sigma}_{2}(W)=\bar{\Sigma}_{1}(-W)$.

Renormalization of spin-$3/2$ sector determines the values $S_{1}(0)$, $S_{2}(0)$ which take
part in cancellation of $1/p^{2}$ singilarities \eqref{eq:app-nosing}.

\subsection{Renormalization of spin-$1/2$ sector}

Let us consider the contraction of spin-$1/2$ sector of propagator with $\theta(a)$, as they
appear in amplitude ($\theta(a)$ is matrix in vertex, see \eqref{eq:interac-lagr})
\begin{equation} \label{eq:MT12}
  Z_{\mu\nu}=\theta_{\mu\alpha}(a)\big( G^{\alpha\beta}\big )^{s=1/2}\theta_{\beta\nu}(a).
\end{equation}
For our purpose it is convenient to write it using the introduced $2\times 2$ matrix
representation \eqref{eq:matr}
\begin{equation}
  Z^{\mu\nu}=T^{\mu\nu}\left[ Z^{+}, Z^{-} \right] = T^{\mu\nu}\Big[
    \theta^{+}(a)G^{+}\theta^{+}(a), \theta^{-}(a)G^{-}\theta^{-}(a) \Big].
\end{equation}
Let us consider in detail the matrix which is accompanied by $\Lambda^{-}$ projector:
\begin{equation}
  Z^{-} = \theta^{-}(a)\ G^{-}\ \theta^{-}(a)= \theta^{-}(a) \left(
    \begin{array}{cc}
      \bar{S}_{3}  & \bar{S}_{7} \\
      \bar{S}_{10} & \bar{S}_{6}
    \end{array} \right)^{-1}
  \theta^{-}(a) .
\end{equation}
The inverse matrix is more transparent:
\begin{equation}
  (Z^{-})^{-1} = \big(\theta^{-}(a)\big)^{-1} \left(
    \begin{array}{cc}
      \bar{S}_{3}  & \bar{S}_{7} \\
      \bar{S}_{10} & \bar{S}_{6}
    \end{array} \right)
  \big(\theta^{-}(a)\big)^{-1}= \big(\theta^{-}(a)\big)^{-1} \Big[ S_{0}^{-} - \Sigma^{-} \Big] \big(\theta^{-}(a)\big)^{-1} .
\end{equation}
Recall that by wave function renormalization we made free propagator to be 2-parametric, as a
result the $\theta(a)$-factor appeared in a vertex. So the self-energy contains this parameter,
as it seen from \eqref{eq:interac-lagr}
\begin{equation}
  \Sigma^{-}(p;a)= \theta^{-}(a)\Sigma^{-}(p;a=0)\theta^{-}(a).
\end{equation}
Matrix $Z$ takes the form
\begin{equation} \label{eq:DPR}
  \big(Z^{-}\big)^{-1} = \big(\theta^{-}(a)\big)^{-1}\Big[S_{0}^{-}-\theta^{-}(a)\Sigma^{-}(p;a=0)\theta^{-}(a)
                        \Big]\big(\theta^{-}(a)\big)^{-1}=
                        \big(\theta^{-}(a)\big)^{-1}S_{0}^{-}\big(\theta^{-}(a)\big)^{-1}-\Sigma^{-}_{a=0}
\end{equation}
and one can see that first term is again the general three-parameter spin-$1/2$ propagator. So
$\big(Z^{-}\big)^{-1}$ is, in fact, the dressed inverse propagator $S^{-}$. We see that after
all the parameter $a$ has been disappered from the self-energy.

\subsubsection{Cancellation of $1/p^{2}$ singularities}
\label{sec:p2-cancel}

First of all require the absence of $1/p^{2}$ singularities in the dressed propagator
\eqref{eq:DPR}. It is convenient to subtract the self-energy at zero (even it is not necessary
for convergence of the integrals)
\begin{equation}
  \Sigma^{-}(W) = \left(
    \begin{array}{cc}
      T_{3}+W T_{4} & \sqrt{3}(\tilde{T}_{8}+W \tilde{T}_{7}) \\
      \sqrt{3}(\tilde{T}_{8}+W \tilde{T}_{7}) & T_{5}-W T_{6}
    \end{array} \right)
  + \Sigma^{-}_{(0)}(W).
\end{equation}
Here $T_{i}$ are subtraction constants of self-energy components in $\hat{p}$-basis%
\footnote{
  Let us suppose here the loop integrals to be convergent, their behaviour is defined by
  form-factor in a vertex, see below.
}%
and $\Sigma_{(0)}(W)$ denotes the integral subtracted at origin,
\begin{equation} \label{eq:selfen}
  \Sigma_{i}(s) = \frac{1}{\pi}\int_{s_{1}}^{\infty}\dfrac{\dd z}{z-s}\,\rho_{i}(z)=T_{i} +
                 \frac{s}{\pi}\int_{s_{1}}^{\infty}\dfrac{\dd z}{z(z-s)}\,\rho_{i}(z)=
                 T_{i} +\Sigma_{(0)i}(s).
\end{equation}
Matrix \eqref{eq:DPR} after it takes the form
\begin{equation}
  \big(Z^{-}\big)^{-1} \equiv S^{-} = S_{0}^{-} - T - \Sigma^{-}_{(0)} = - T - \Sigma^{-}_{(0)}.
\end{equation}
We can see that both matrices $S_{0}^{-}$ and $T$ have the same structure (see \eqref{eq:mostl}
and \eqref{eq:app-nosing1}) and it leads only to changing of parameters. So we can omit
$S_{0}^{-}$ in previous formula. After this it is easy to write down reparametrized expression
which is free from $1/p^{2}$ singularities for fixed spin-$3/2$ sector
\begin{equation} \label{eq:Sfree}
  S^{-}= - \left(
    \begin{array}{ccc}
      2S_{1}(0)+3T_{5} & & \sqrt{3}(-S_{1}(0)-T_{5}) \\
      \sqrt{3}(-S_{1}(0)-T_{5})& & T_{5}
    \end{array} \right)
  - W \left(
    \begin{array}{ccc}
      2S_{2}(0)+3T_{6}-6\tilde{T}_{7} & & \sqrt{3}\tilde{T}_{7} \\
      \sqrt{3}\tilde{T}_{7} & & - T_{6}
    \end{array} \right)
  - \Sigma^{-}_{(0)}.
\end{equation}
We used here the relations \eqref{eq:app-nosing}, \eqref{eq:app-nosing1} between coefficients.
This equation contains three parameters $T_{5}$, $T_{6}$, and $\tilde{T}_{7}$, while $S_{1}(0)$,
$S_{2}(0)$ are fixed after renormalization of the spin-$3/2$ sector.

\subsubsection{Behaviour at infinity}
\label{sec:behav-at-infin}

Let us consider determinant of \eqref{eq:Sfree} at $W\to\infty$.
\begin{equation}
  \Delta^{-}=\det(S^{-})= W^{2} d_{2} + W d_{1} + d_{0} \quad \text{at} \quad W\to\infty .
\end{equation}
Note that subtracted loops $\Sigma_{(0)i}(W^2)$ tends to some constants $\Sigma_{(0)i}(\infty)$
in this limit. It is convenient to introduce new notations
\begin{equation}
  \tau_{i} = T_{i} + \Sigma_{(0)i}(\infty).
\end{equation}
There are appeared some natural combinations of parameters, containing $S_{1}(0)$, $S_{2}(0)$:
\begin{equation}
  \begin{split}
    C_{1}&= S_{1}(0)- \Sigma_{(0)5}(\infty), \\
    C_{2}&=2S_{2}(0)+ \Sigma_{(0)4}(\infty)-3\Sigma_{(0)6}(\infty)+6\Sigma_{(0)7}(\infty), \\
    C_{3}&=2S_{1}(0)+ \Sigma_{(0)3}(\infty)-3\Sigma_{(0)5}(\infty).
  \end{split}
\end{equation}
Let us require $\Delta^{-}\to\const$ at $W\to\infty$. It gives
\begin{equation} \label{eq:eqD}
  \begin{split}
    d_{2}&=-C_{2} \tau_{6} -3(\tau_{6}-\tau_{7})^{2} =0,         \\
    d_{1}&= C_{2} \tau_{5}-C_{3} \tau_{6} + 6 C_{1} \tau_{7} =0, \\
    d_{0}&=-3C_{1}^{2}+\tau_{5}(C_{3}-6C_{1}).
  \end{split}
\end{equation}
These two equations are solved easily if to introduce the sum and difference instead of $\tau_{6},$
$\tau_{7}$: $\delta=\tau_{6}-\tau_{7}$, $\sigma=\tau_{6}+\tau_{7}$. Solving \eqref{eq:eqD}, we
express all parameters through the one free parameter $\delta$.
\begin{equation}
  \sigma=-\frac{6\delta^{2}}{C_{2}}-\delta,\quad
  \tau_{5}=\frac{3\delta}{C_{2}^{2}}\Big[-\delta C_{3} + 2C_{2}(3\delta+C_{2})\Big].
\end{equation}
Final expressions for $\tau_{i}$ are (we introduced here notation $\delta=-A-1$ for historical reasons)
\begin{equation} \label{eq:taus}
  \begin{split}
    \tau_{3}  &= \dfrac{(C_{2}-3A-3)(3C_{3}A-18C_{1}A+3C_{3}-18C_{1} + C_{2}C_{3})}{C_{2}^{2}}, \\
    \tau_{4}  &= \dfrac{(3A+3-C_{2})^{2}}{C_{2}}, \\
    \tau_{5}  &= -3\dfrac{(A+1)(2C_{1} C_{2}-6C_{1}A-6C_{1}+C_{3}A+C_{3})}{C_{2}^{2}}, \\
    \tau_{6}  &= -3\dfrac{(A+1)^{2}}{C_{2}}, \\
    \tau_{7}  &= \dfrac{(A+1)(C_{2}-3A-3)}{C_{2}}, \\
    \tau_{8}  &= \dfrac{(3C_{3}-18C_{1})(A+1)^{2}+6C_{1}C_{2}(A+1)+C_{1}C_{2}^{2}}{C_{2}^{2}}, \\
    \tau_{9}  &= \tau_{7}, \\
    \tau_{10} &=-\tau_{8}.
  \end{split}
\end{equation}
Recall that $\tau_{i}$ are subtraction constants at infinity, so renormalized propagator may be
written as
\eqref{eq:Sfree}
\begin{equation} \label{eq:reno}
  S^{-}= - \left(
    \begin{array}{ccc}
      \tau_{3}+\Sigma_{3}(s) & &   \sqrt{3}(\tau_{8}+\Sigma_{8}(s)) \\
        \sqrt{3}(\tau_{8}+\Sigma_{8}(s))    & & \tau_{5}+\Sigma_{5}(s)
    \end{array} \right)
  - W \left(
    \begin{array}{ccc}
           \tau_{4}+\Sigma_{4}(s)     & & \sqrt{3}(\tau_{7}+\Sigma_{7}(s)) \\
      \sqrt{3}(\tau_{7}+\Sigma_{7}(s))   & &    - \tau_{6} - \Sigma_{6}(s)
    \end{array} \right),
\end{equation}
where $\Sigma_{i}(s)$ are non-subtracted integrals \eqref{eq:selfen} and $\tau_{i}$ are defined
by \eqref{eq:taus}.

If in renormalized propagator \eqref{eq:reno} to turn off interaction ($\Sigma_{i}(s)=0$) and to put
$C_{i}$ to its bare values
\begin{equation}
  C_{1}^{0}=-M, \;  C_{2}^{0}=2, \; C_{3}^{0}=-2M
\end{equation}
then we come back to famous one-parameter free inverse propagator \cite{MC1956} with unphysical
spin-$1/2$ sector
\begin{equation} \label{eq:S0}
  \begin{split}
    S^{-}_{0}&=\left(
      \begin{array}{cc}
            M(3A+1)(3A+2)      & -\sqrt{3}M(3A^{2}+3A+1) \\
        -\sqrt{3}M(3A^{2}+3A+1) &         3MA(A+1)
      \end{array} \right) \\[1mm]
    &+ \frac{W}{2} \left(
      \begin{array}{cc}
            -(3A+1)^{2}      & \sqrt{3}(A+1)(3A+1) \\
        \sqrt{3}(A+1)(3A+1) &     - 3(A+1)^{2}
      \end{array} \right)
  \end{split}
\end{equation}
Determinant of this matrix is equal to constant $\det(S^{-})=-2M^{2}(2A+1)^{2}$, so there is no
poles in spin-$1/2$ sector.



\section{Application to $\pi\particle{N}$ scattering}

As a first step we will use the obtained renormalized \RS propagator for description of total
cross-section $\pi^{+}\particle{p}\to\pi^{+}\particle{p}$ in vicinity of $\Delta^{++}(1232)$
resonance. The discussed problem of non-leading spin terms in \RSf, on the other hand, is a
problem of exact form of resonance curve. As we will see below, the high accuracy $\pi\particle{N}$ data
allow to "feel" these non-leading spin terms due to interference with main spin-$3/2$
contribution even in total cross-section.

\subsection{Amplitude $\pi\particle{N}\to \Delta \to \pi\particle{N}$}
\label{sec:app-pND-amp}

Standard $\pi\particle{N}\Delta$ interaction lagrangian is of the form%
\footnote{%
  We are interested only in isospin $I=3/2$ so we omit here isotopical indices.
}%
\begin{equation}
  \LagrInt=g_{\pi\particle{N}\Delta}\bar{\Psi}\vphantom{\Psi}_{\mu}(x)\theta^{\mu\nu}(a)\Psi(x)\cdot\partial_{\nu}\phi(x) + \hc
\end{equation}
Resonance contribution:
\begin{figure}[h]
  \centering
  \includegraphics*[width=4.0cm]{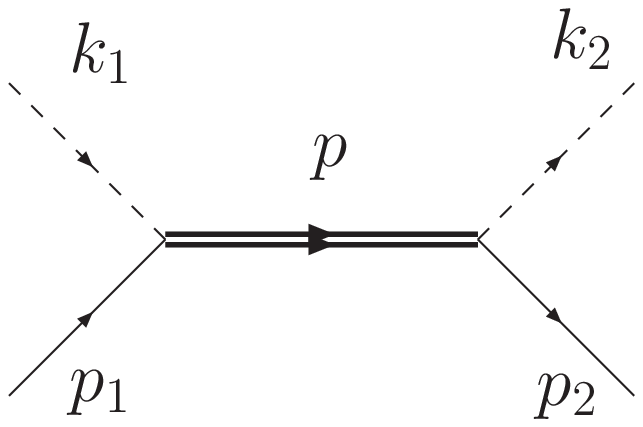}
\end{figure}

\begin{equation} \label{eq:app-R32}
  \matel=g_{\pi\particle{N}\Delta}^{2}\cdot \bar{u}(p_{2})k_{2}^{\mu}\theta_{\mu\alpha}(a)G^{\alpha\beta}(p)\theta_{\beta\nu}(a)k_{1}^{\nu} u(p_{1}).
\end{equation}
Here $G^{\alpha\beta}(p)$ is dressed propagator of \RSf,
$\theta^{\mu\alpha}(a)=g^{\mu\alpha}+a\gamma^{\mu} \gamma^{\alpha}$. As it was seen above
\eqref{eq:MT12} and \eqref{eq:DPR} the $a$ parameter has been disappeared from the amplitude. So
we can put $a=0$ below.

Let us write down the matrix between spinors in \eqref{eq:app-R32}.

\subsubsection{Spin-3/2 contribution}
\label{sec:app-amp-32}

\begin{equation} \label{eq:app-32}
  \begin{split}
    R^{s=3/2} &= k_{2\mu}\big(G^{\mu\nu}\big)^{s=3/2}k_{1\nu} = k_{2\mu}\big(\operP^{\mu\nu}_{1}\bar{G}_{1}+
               \operP^{\mu\nu}_{2}\bar{G}_{2}\big) k_{1\nu}= \\
             &= \Lambda^{+} \biggl[ -\bar{G}_{1} \vect{p}^{2} \cos{\theta} - \bar{G}_{2}
               \dfrac{[(W-m)^{2} - \mu^{2}]^{2}}{12 W^{2}} \biggr] + \\
             &+ \Lambda^{-} \biggl[ -\bar{G}_{2} \vect{p}^{2} \cos{\theta} - \bar{G}_{1} \dfrac{[(W+m)^{2}
              - \mu^{2}]^{2}}{12 W^{2}} \biggr].
  \end{split}
\end{equation}
Here $\vect{p}$ is a spatial momentum of $\pi\particle{N}$ pair in
c.m.s. $\vect{p}^{2}=\lambda(s,\mN^{2},\mpi^{2})/4s$.

\subsubsection{Spin-1/2 contribution}
\label{sec:app-amp-12}

\begin{equation} \label{eq:app-12}
  \begin{split}
    R^{s=1/2} &= k_{2}^{\mu}\theta_{\mu\alpha}(0)\big(G^{\alpha\beta}\big)^{s=1/2}\theta_{\beta\nu}(0)k_{1}^{\nu}=
    k_{2}^{\mu}\theta_{\mu\alpha}(0)\Bigl(\sum_{i=3}^{10}\operP^{\alpha\beta}_{i}\bar{G}^{i}\Bigr)\theta_{\beta\nu}(0)k_{1}^{\nu} = \\
    &=
      \begin{pmatrix}
        k_{2}\theta n_{1} \Lambda^{+}, & k_{2}\theta n_{2} \Lambda^{+}
      \end{pmatrix}
      \begin{pmatrix}
        \bar{G}_{4} & \bar{G}_{8} \\
        \bar{G}_{9} & \bar{G}_{5}
      \end{pmatrix}
      \begin{pmatrix}
        \Lambda^{+} n_{1} \theta k_{1} \\
        \Lambda^{+} n_{2} \theta k_{1}
      \end{pmatrix}
    + \\
    &+
      \begin{pmatrix}
        k_{2}\theta n_{1} \Lambda^{-}, &  k_{2}\theta n_{2} \Lambda^{-}
      \end{pmatrix}
      \begin{pmatrix}
        \bar{G}_{3}  & \bar{G}_{7} \\
        \bar{G}_{10} & \bar{G}_{6}
      \end{pmatrix}
      \begin{pmatrix}
        \Lambda^{-} n_{1} \theta k_{1} \\
        \Lambda^{-} n_{2} \theta k_{1}
      \end{pmatrix}.
  \end{split}
\end{equation}
Projection operators and spinors turn the components of "vectors" into unit matrix $4\times4$.

\begin{equation}
  \begin{split}
    R^{s=1/2} &= \Lambda^{+} \cdot
    \begin{pmatrix}
      a_{1}(W), & a_{2}(W)
    \end{pmatrix}
    \begin{pmatrix}
      \bar{G}_{4} & \bar{G}_{8} \\
      \bar{G}_{9} & \bar{G}_{5}
    \end{pmatrix}
    \begin{pmatrix}
      a_{1}(W) \\
      a_{2}(W)
    \end{pmatrix}
    + \\
    &+ \Lambda^{-} \cdot
    \begin{pmatrix}
      a_{1}(-W), & a_{2}(-W)
    \end{pmatrix}
    \begin{pmatrix}
      \bar{G}_{3}  & \bar{G}_{7} \\
      \bar{G}_{10} & \bar{G}_{6}
    \end{pmatrix}
    \begin{pmatrix}
      a_{1}(-W) \\
      a_{2}(-W)
    \end{pmatrix}.
  \end{split}
\end{equation}
Here
\begin{equation}
  \begin{split}
    a_{1}(W) &= \dfrac{1}{2\sqrt{3}W}\big[ (W-\mN)^{2}-\mpi^{2}\big],\\
    a_{2}(W) &= \dfrac{1}{2W}\big[ W^{2}-\mN^{2}+\mpi^{2}\big].
  \end{split}
\end{equation}
In more detail:
\begin{multline}
  R^{s=1/2} = \Lambda^{+} \Big[ a^{2}_{1}(W) \bar{G}_{4} + a_{1}(W)a_{2}(W)
  \bar{G}_{8} + a_{1}(W)a_{2}(W) \bar{G}_{9} + a^{2}_{2}(W) \bar{G}_{5}\Big] + \\
  + \Lambda^{-} \Big[ a^{2}_{1}(-W) \bar{G}_{3} - a_{1}(-W)a_{2}(-W) \bar{G}_{7} -
  a_{1}(-W)a_{2}(-W) \bar{G}_{10} + a^{2}_{2}(-W) \bar{G}_{6}\Big].
\end{multline}

\subsection{Self-energy}

The one-loop self-energy contribution is
\begin{equation}
  \Sigma^{\mu\nu}(p)=-\imath g_{\pi\particle{N}\Delta}^{2} \int\frac{\dd^{4}k}{(2\pi)^{4}}\,\theta^{\mu\rho}(0)
                   k_{\rho}\,\frac{1}{\hat{p}+\hat{k}-\mN}\,k_{\lambda}\theta^{\lambda\nu}(0)\,\frac{1}{k^{2}-\mpi^{2}}.
\end{equation}
Let us calculate the discontinuity of loop contribution in $\hat{p}$-basis.
\begin{equation} \label{eq:disc}
  \begin{array}{ll}
    \begin{array}{l}
      \Delta \Sigma_{1} = -\imath\dfrac{g^{2}I_{0}}{(2\pi)^{2}} \dfrac{\mN}{12s}\lambda(s,\mN^{2},\mpi^{2}),  \\
      \Delta \Sigma_{2} = -\imath\dfrac{g^{2}I_{0}}{(2\pi)^{2}} \dfrac{1}{24s^{2}}(s+\mN^{2}-\mpi^{2})\lambda ,
    \end{array}
   &\begin{array}{l}
      \Delta \Sigma_{3} = -\imath\dfrac{g^{2}I_{0}}{(2\pi)^{2}} \dfrac{\mN}{12s}\lambda ,  \\
      \Delta \Sigma_{4} = -\imath\dfrac{g^{2}I_{0}}{(2\pi)^{2}} \dfrac{1}{24s^{2}}(s+\mN^{2}-\mpi^{2})\lambda ,
    \end{array}\\
    \begin{array}{l}
      \Delta \Sigma_{5} = \imath\dfrac{g^{2}I_{0}}{(2\pi)^{2}} \dfrac{\mN}{4s}(s-\mN^{2}+\mpi^{2})^{2},  \\
      \Delta \Sigma_{6} = \imath\dfrac{g^{2}I_{0}}{(2\pi)^{2}} \dfrac{1}{8s^{2}}(s+\mN^{2}-\mpi^{2})(s-\mN^{2}+\mpi^{2})^{2},
    \end{array}
   &\begin{array}{l}
      \Delta \Sigma_{7} = \imath\dfrac{g^{2}I_{0}}{(2\pi)^{2}} \sqrt{\dfrac{3}{s}}\cdot\dfrac{1}{24s}(s-\mN^{2}+\mpi^{2})\lambda, \\
      \Delta \Sigma_{8} =  0 ,              \\
      \Delta \Sigma_{9} = \Delta \Sigma_{7}, \\
      \Delta \Sigma_{10}= 0 .
    \end{array}
  \end{array}
\end{equation}
Here $\lambda(a,b,c)=(a-b-c)^{2}-4bc$, arguments of $\lambda$ are shown in the first
expression. $I_{0}$ is the base integral
\begin{equation} \label{eq:ibase}
  \begin{split}
    I_{0} &= \int\dd^{4}k\delta\big(k^{2}-\mpi^{2}\big)\delta\big((p+k)^{2}-\mN^{2}\big)=
            \theta\big(p^{2}-(\mN+\mpi)^{2}\big)\frac{\pi}{2}\sqrt{\frac{\lambda\big(p^{2},\mN^{2},\mpi^{2}\big)}{\big(p^{2}\big)^{2}}}.
  \end{split}
\end{equation}
Let us write down also the imaginary parts of first components in $\Lambda$-basis:
\begin{equation}
  \begin{split}
    \Im \bar{\Sigma}_{1} &= \Im \Sigma_{1}(W^{2})+W \Im \Sigma_{2}(W^{2})=-\frac{g^{2}}{16\pi}\,
    \frac{\lambda^{3/2}}{24W^{5}}\Big[(W+\mN)^{2}-\mpi^{2}\Big]= -\Gamma(W)/2  \\
    \Im \bar{\Sigma}_{2} &= \Im \Sigma_{1}(W^{2})-W \Im \Sigma_{2}(W^{2})=\frac{g^{2}}{16\pi}\,
    \frac{\lambda^{3/2}}{24W^{5}}\Big[(W-\mN)^{2}-\mpi^{2}\Big].
  \end{split}
\end{equation}
And last: let us introduce $W$-dependent form-factor into coupling constant (also known as
centrifugal form-factor or Blatt--Weisskopf form-factor, see e.g. \cite{Nagels1979,Cut1979}).
We will make it in a simplest way to ensure the convergence of loop integrals
\begin{equation}\label{FF}
  g \to g\cdot F(W)= g\cdot \frac{M^{2}+\sL}{W^{2}+\sL}.
\end{equation}
Such modification allows to perform the renormalization procedure in the spin-$1/2$ sector which was
described above.

\subsection{Fit of total cross-section $\pi^{+}\particle{p}\to \Delta^{++}(1232)\to \pi^{+}\particle{p}$}

Let us consider the data $\pi^{+}\particle{p}\to\pi^{+}\particle{p}$ \cite{Pedroni1978} which
have the best statistics in the region of $\Delta(1232)$ resonance. We can use the obtained
renormalized R.--S. propagator for description of total cross-section.  Results of fit in energy
region from the threshold up to $1.32\;\GeV$ are presented at Fig.~\ref{fig:CS}. Note that use of
dressed propagator of R.--S. field with all spin components leads to good quality of
description. The best-fit parameters are:
\begin{equation}\label{eq:best}
  \begin{array}{cc}
    \begin{array}{ll}
    M_{\Delta}      &= 1232.0 \pm 0.2\;\; \MeV   \\
    \Gamma_{\Delta} &= 11   3 \pm 1  \;\; \MeV
    \end{array}
    &
    \begin{array}{rl}
    A             &= -0.576 \pm 0.014             \\
    \sL           &= -0.607 \pm 0.012\;\; \GeV^{2} \\
    \chi^{2}/DOF   &= 0.99.
    \end{array}
  \end{array}
\end{equation}
In Table~\ref{tab:comp} one can see comparison of different variants of fit, shown at
Fig.~\ref{fig:CS}. Evidently that account of spin-$1/2$ terms in dressed propagator improves
essentially quality of description. For comparison at Fig.~\ref{fig:CS} one can see also the
curve, corresponding to naive Breit-Wigner contribution with energy-independent mass and width.

Looking at the best-fit parameters \eqref{eq:best} we observe at first sight unexpected fact:
the parameter $\sL$ is negative, so we have pole in the vertex formfactor \eqref{FF} not far
from $\pi N$ threshold. We suppose that appearance of this pole imitates the one-nucleon
contribution. Of course, simplest parameterization \eqref{FF} does not reproduce correctly the
one-nucleon term but data "feel" the presence of rapidly changing under-threshold
contribution. In principle it's possible to improve the model of formfactor to reproduce the
one-nucleon term. Recall that there exist an approach based on low-energy dynamical dispersion
equations (see, e.g. \cite{SSM1969}), which account correctly the cross-channels contributions. So
we should make our amplitude to be consistent with these dynamical equations. But as a first
step we restrict ourselves by rough recipe \eqref{FF}.

After fixing free parameters by \eqref{eq:best}, we can calculate partial waves of isospin-$3/2$
with use of our dressed propagator.  Our amplitude contains four partial waves $P_{33}$,
$D_{33}$, $S_{31}$, $P_{31}$, satisfying the elastic unitary condition $\Im f= \abs{f}^{2}$. It turns
out that the so obtained partial waves are in reasonable agreement with results of the partial
wave analysis \cite{Arndt2006}, as it seen from Figs.~\ref{fig:PW},\ref{fig:PW2}.

Note that at $A\sim -0.5$ in our amplitudes takes place some changing of behaviour. This is
rather natural if to remember that $A=-0.5$ is the singular value for free propagator.



\section{Conclusions}
\label{sec:concls}

Thus we found the renormalization procedure for dressed propagator of Rarita-Schwinger field,
which remains the spin-$1/2$ degrees of freedom to be unphysical. This fact is rather
non-trival, as it seen from discussion in \cite{PT1999}. We think that
the studying of dressed propagator is more adequate method then investigation of equations of
motion and additional constrains.

Spin-$3/2$ contribution after renormalization takes usual resonance form.  The obtained
renormalized propagator in $s=1/2$ sector has one free parameter $A$ and is a straight
generalization of the standard free propagator, suggested by Moldauer and Case \cite{MC1956}.

We found that the obtained dressed propagator describes well the total cross-section
$\pi^{+}\particle{p}\to \pi^{+}\particle{p}$ in vicinity of $\Delta^{++}(1232)$
resonance. Moreover, it allows to describe the partial waves also. If to say more exactly, our
amplitude describes well $J=3/2$ waves $P_{33}$ and $D_{33}$. As for $J=1/2$ partial waves, it
describes reasonably the smooth non-resonant contributions in these waves.

So we can conslude that the concept of effective multicomponent quantum field is really working
thing and it may be useful in phenomenology.


\section{Acknowledgments}

We thank A.M. Moiseeva for participation at the beginning of this
work. This work was supported in part by RFBR grant No
05-02-17722.



\appendix

\section{Decomposition of spin-tensor}
\label{sec:app-sp-decomp}

Propagator or self-energy of the \RSf has two spinor and two vector indices and depends
on momentum $p$. We will denote such object as $S^{\mu\nu}(p)$, omitting spinor indices, and
will call it shortly as a spin-tensor. In our consideration we need to use different bases for
this object.

\subsection{$\gamma$-basis}
\label{sec:app-gb}

Most evident is a $\gamma$-matrix decomposition. It's easy to write down all possible
$\gamma$-matrix structures with two vector indices. Altogether there are 10 terms in
decomposition of spin-tensor, if parity is conserved.
\begin{equation} \label{eq:app-eq1}
  \begin{split}
    S^{\mu\nu}(p) &= g^{\mu\nu}\cdot s_{1}+p^{\mu}p^{\nu}\cdot s_{2} +\hat{p}p^{\mu}p^{\nu}\cdot
                    s_{3}+\hat{p}g^{\mu\nu}\cdot s_{4}+p^{\mu}\gamma^{\nu}\cdot s_{5}+\gamma^{\mu}p^{\nu}\cdot s_{6}+ \\
                 &+\sigma^{\mu\nu}\cdot s_{7}+\sigma^{\mu\lambda}p_{\lambda}p^{\nu}\cdot s_{8}+\sigma^{\nu\lambda}
                 p_{\lambda}p^{\mu}\cdot s_{9}+\gamma_{\lambda}\gamma^{5}\imath\epsilon^{\lambda\mu\nu\rho}p_{\rho}\cdot s_{10}.
  \end{split}
\end{equation}
Here $s_{i}(p^{2})$ are the Lorentz-invariant coefficients and
$\sigma^{\mu\nu}=\frac{1}{2}[\gamma^{\mu},\gamma^{\nu}]$.

This is a good starting point of any consideration, since this basis is complete, nonsingular
and free of kinematical constrains.  But this basis is not convenient at multiplication (e.g. in
Dyson summation) because elements of basis are not orthogonal to each other.

\subsection{$\hat{p}$-basis}
\label{sec:app-pb}

There is another basis (see e.g. \cite{PS1995}) for $S^{\mu\nu}$, which we call as
$\hat{p}$-basis. Decomposition of any spin-tensor in this basis has the form
\begin{multline} \label{eq:app-eq2}
  S^{\mu\nu}(p)=(S_{1}+\hat{p}S_{2})\big(\operP^{3/2}\big)^{\mu\nu}+(S_{3}+
  \hat{p}S_{4})\big(\operP^{1/2}_{11}\big)^{\mu\nu}
  +(S_{5}+\hat{p}S_{6})\big(\operP^{1/2}_{22}\big)^{\mu\nu}+ \\
  +(S_{7}+\hat{p}S_{8})\big(\operP^{1/2}_{21}\big)^{\mu\nu}+
  (S_{9}+\hat{p}S_{10})\big(\operP^{1/2}_{12}\big)^{\mu\nu},
\end{multline}
where appeared the well-known tensor operators
\begin{equation} \label{eq:app-eq3}
  \begin{split}
    (\operP^{3/2})^{\mu\nu}      &=g^{\mu\nu}-n_{1}^{\mu}n_{1}^{\nu}-n_{2}^{\mu}n_{2}^{\nu}\\
    (\operP^{1/2}_{11})^{\mu\nu} &=n_{1}^{\mu}n_{1}^{\nu}, \\
    (\operP^{1/2}_{22})^{\mu\nu} &=n_{2}^{\mu}n_{2}^{\nu}, \\
    (\operP^{1/2}_{21})^{\mu\nu} &=n_{1}^{\mu}n_{2}^{\nu}, \\
    (\operP^{1/2}_{12})^{\mu\nu} &=n_{2}^{\mu}n_{1}^{\nu}.
  \end{split}
\end{equation}
which are written here in a non-standard form. Here we introduced the unit "vectors"
\begin{equation}
  n_{1}^{\mu}=\pi^{\mu}/\pi^{2},       \quad
  n_{2}^{\mu}=p^{\mu}/\sqrt{p^2}I_{4}, \quad
  (n_{i}\cdot n_{j})=\delta_{ij}I_{4},
\end{equation}
where "vector" $\pi^{\mu}$ is
\begin{equation*}
  \pi^{\mu}=\frac{1}{3p^{2}}(-p^{\mu}+\gamma^{\mu}\hat{p})\hat{p}
\end{equation*}
with the following properties:
\begin{equation}
  (\pi p)=0,\quad (\gamma\pi)=(\pi\gamma)=1,\quad (\pi\pi)=\frac{1}{3},
  \quad \hat{p}\pi^{\mu}=-\pi^{\mu}\hat{p}.
\end{equation}

\subsection{$\Lambda$-basis}
\label{sec:app-lb}

The most convenient at multiplication basis is build by combining the $\operP^{i}_{\mu\nu}$
operators \eqref{eq:app-eq3} and the off-shell projection operators $\Lambda^{\pm}$
\begin{equation}  \label{eq:app-project}
  \Lambda^{\pm}=\frac{1}{2} \Big(1 \pm \frac{\hat{p}}{\sqrt{p^{2}}} \Big),
\end{equation}
where we assume $\sqrt{p^{2}}$ to be the first branch of analytical function. Ten elements of
this basis look as
\begin{flalign} \label{eq:app-eq6}
  \mathcal{P}_{1} &= \Lambda^{+}\mathcal{P}^{3/2},\,&
  \mathcal{P}_{3} &= \Lambda^{+}\mathcal{P}^{1/2}_{11},\,&
  \mathcal{P}_{5} &= \Lambda^{+}\mathcal{P}^{1/2}_{22},\,&
  \mathcal{P}_{7} &= \Lambda^{+}\mathcal{P}^{1/2}_{21},\,&
  \mathcal{P}_{9} &= \Lambda^{+}\mathcal{P}^{1/2}_{12},\notag\\
  \mathcal{P}_{2} &= \Lambda^{-}\mathcal{P}^{3/2},\,&
  \mathcal{P}_{4} &= \Lambda^{-}\mathcal{P}^{1/2}_{11},\,&
  \mathcal{P}_{6} &= \Lambda^{-}\mathcal{P}^{1/2}_{22},\,&
  \mathcal{P}_{8} &= \Lambda^{-}\mathcal{P}^{1/2}_{21},\,&
  \mathcal{P}_{10}&= \Lambda^{-}\mathcal{P}^{1/2}_{12},
\end{flalign}
where tensor indices are omitted. Decomposition in this basis:
\begin{equation} \label{eq:app-lambda}
  S^{\mu\nu}(p)= \sum_{A=1}^{10}\bar{S}^{A} \operP_{A}^{\mu\nu}.
\end{equation}
The multiplication table of elements of the basis is in Table~\ref{tab:tt}

Coefficients of $S^{\mu\nu}$ in $\hat{p}$- and $\gamma$-bases are related by
\begin{gather} \label{eq:app-eq4}
  \begin{array}{ll}
    s_{1}=\dfrac{1}{3}(2 S_{1} + S_{3}), \quad\quad &
    s_{2}=\dfrac{1}{3W^{2}}( -2 S_{1} - S_{3} + 3 S_{5} ),\\
    s_{3}=\dfrac{1}{3W^{2}}\big( -2S_{2} - S_{4} + 3S_{6} - \dfrac{\sqrt{3}}{W} (S_{7}+S_{9}) \big), \quad\quad &
    s_{4}=\dfrac{1}{3}(2 S_{2} + S_{4}),\\
    s_{5}=\dfrac{1}{\sqrt{3}W}S_{9},  \quad\quad &
    s_{6}=\dfrac{1}{\sqrt{3}W}S_{7},              \\
    s_{7}=\dfrac{1}{3}( - S_{1} + S_{3}),  \quad\quad &
    s_{8}=\dfrac{1}{3W^{2}}\big( S_{1} - S_{3} - \sqrt{3} W S_{8}\big),\\
    s_{9}=\dfrac{1}{3W^{2}}\big( -S_{1} + S_{3} - \sqrt{3} W S_{10}\big),  \quad\quad &
    s_{10}=\dfrac{1}{3}( - S_{2} + S_{4}).
  \end{array}
\end{gather}

Reversed relations:
\begin{gather} \label{eq:app-eq5}
  \begin{array}{ll}
    S_{1}=s_{1}-s_{7},                 \quad\quad &
    S_{2}=s_{4}-s_{10},                \\
    S_{3}=s_{1}+2s_{7},                \quad\quad &
    S_{4}=s_{4}+2s_{10},               \\
    S_{5}=s_{1}+W^{2}s_{2},            \quad\quad &
    S_{6}=W^{2}s_{3}+s_{4}+s_{5}+s_{6},  \\
    S_{7}=\sqrt{3}Ws_{6},             \quad\quad &
    S_{8}=-\dfrac{\sqrt{3}}{W}( s_{7} + W^{2} s_{8}),\\
    S_{9}=\sqrt{3}Ws_{6},             \quad\quad &
    S_{10}=\dfrac{\sqrt{3}}{W}( s_{7}+ W^{2} s_{8}).
  \end{array}
\end{gather}
Transition from $\hat{p}$- to $\Lambda$-basis:
\begin{flalign} \label{eq:app-eq7}
   \bar{S}_{1} &= S_{1}+W S_{2},
  &\bar{S}_{3} &= S_{3}+W S_{4},
  &\bar{S}_{5} &= S_{5}+W S_{6},
  &\bar{S}_{7} &= S_{7}+W S_{8},
  &\bar{S}_{9} &= S_{9}+W S_{10},\notag\\
   \bar{S}_{2} &= S_{1}-W S_{2},
  &\bar{S}_{4} &= S_{3}-W S_{4},
  &\bar{S}_{6} &= S_{5}-W S_{6},
  &\bar{S}_{8} &= S_{7}-W S_{8},
  &\bar{S}_{10}&=S_{9}-W S_{10}.
\end{flalign}
Let us note that $\hat{p}$- and $\Lambda$-bases are singular at point $p^{2}=0$. As for branch
point $\sqrt{p^{2}}$ appearing in different terms, it is canceled in total expression. But poles
$1/p^{2}$ don't cancel automatically, if we work in $\hat{p}$- or $\Lambda$-basis. First of all,
one can see from \eqref{eq:app-eq5} that the $S_{7}-S_{10}$ should have kinematical $\sqrt{p^{2}}$
factors:
\begin{equation} \label{eq:app-kinfac}
  S_{7}=\sqrt{3}W\tilde{S}_{7},\quad
  S_{8}=\dfrac{\sqrt{3}}{W}\tilde{S}_{8},\quad
  S_{9}=\sqrt{3}W\tilde{S}_{9},\quad
  S_{10}=\dfrac{\sqrt{3}}{W}\tilde{S}_{10},
\end{equation}
and $\tilde{S}_{i}$ don't have branch point at origin. After it we see from \eqref{eq:app-eq4}
conditions of absence of $1/p^{2}$ poles:
\begin{equation} \label{eq:app-nosing}
  \begin{split}
    2S_{1}(0)+S_{3}(0)-3S_{5}(0)&=0,                                     \\
    S_{1}(0)-S_{3}(0)-3\tilde{S}_{8}(0)&=0,                              \\
    2S_{2}(0)+S_{4}(0)-3S_{6}(0)+3(\tilde{S}_{7}(0)+\tilde{S}_{9}(0))&=0, \\
    \tilde{S}_{8}(0)+\tilde{S}_{10}(0)&=0.
  \end{split}
\end{equation}
It is convenient to solve these relations in such manner:
\begin{equation} \label{eq:app-nosing1}
  \begin{split}
    S_{3}&=-2S_{1}(0)+3S_{5}(0),                                        \\
    \tilde{S}_{8}(0)&=S_{1}(0)-S_{5}(0),                                \\
    S_{4}(0)&=-2S_{2}(0)+3S_{6}(0)-3(\tilde{S}_{7}(0)+\tilde{S}_{9}(0)), \\
    \tilde{S}_{10}(0)&=-\tilde{S}_{8}(0).
  \end{split}
\end{equation}








\newpage

\begin{table}[ht]
  \caption{\label{tab:tt} Multiplicative properties of the $\Lambda$-basis}
  \begin{tabular}{X|XXXXXXXXXX}
    \qquad & $\mathcal{P}_{1}$ & $\mathcal{P}_{2}$ & $\mathcal{P}_{3}$ & $\mathcal{P}_{4}$ & $\mathcal{P}_{5}$ & $\mathcal{P}_{6}$ & $\mathcal{P}_{7}$ & $\mathcal{P}_{8}$ & $\mathcal{P}_{9}$ & $\mathcal{P}_{10}$ \\ \hline
    $\mathcal{P}_{1}$    & $\mathcal{P}_{1}$ & 0&0&0&0&0&0&0&0&0              \\
    $\mathcal{P}_{2}$    &0&$\mathcal{P}_{2}$&0&0&0&0&0&0&0&0                 \\
    $\mathcal{P}_{3}$    &0&0&$\mathcal{P}_{3}$&0&0&0&$\mathcal{P}_{7}$&0&0&0  \\
    $\mathcal{P}_{4}$    &0&0&0&$\mathcal{P}_{4}$&0&0&0&$\mathcal{P}_{8}$&0&0  \\
    $\mathcal{P}_{5}$    &0&0&0&0&$\mathcal{P}_{5}$&0&0&0&$\mathcal{P}_{9}$&0  \\
    $\mathcal{P}_{6}$    &0&0&0&0&0&$\mathcal{P}_{6}$&0&0&0&$\mathcal{P}_{10}$ \\
    $\mathcal{P}_{7}$    &0&0&0&0&0&$\mathcal{P}_{7}$&0&0&0&$\mathcal{P}_{3}$  \\
    $\mathcal{P}_{8}$    &0&0&0&0&$\mathcal{P}_{8}$&0&0&0&$\mathcal{P}_{4}$&0  \\
    $\mathcal{P}_{9}$    &0&0&0&$\mathcal{P}_{9}$&0&0&0&$\mathcal{P}_{5}$&0&0  \\
    $\mathcal{P}_{10}$ &0&0&$\mathcal{P}_{10}$&0&0&0&$\mathcal{P}_{6}$&0&0&0   \\
  \end{tabular}
\end{table}

\begin{table}[h]
  \centering
  \caption{\label{tab:comp} Comparison of different models at fitting
    of cross section shown at Fig.~\ref{fig:CS}}
  \begin{tabular}{|p{6cm}|c|}
    \hline
    \hspace{2cm} Model & $\chi^{2}/DOF$ \\
    \hline\hline
    Simplest Breit-Wigner with energy independent mass and width & $215$ \\
    \hline
    Dressed spin-$3/2$ components & $5.6$ \\
    \hline
    Pressed propagator of \RSf with all components & $0.99$ \\
    \hline
  \end{tabular}
\end{table}

\newpage

\begin{figure}[h]
  \includegraphics*[width=9cm]{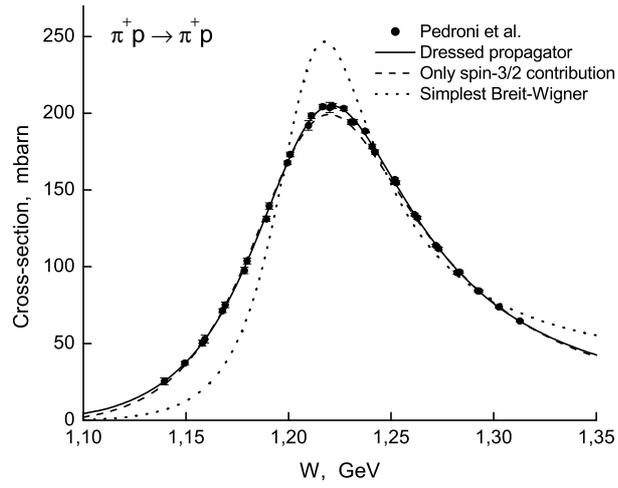}
  \caption{\label{fig:CS} Results of fit of data \cite{Pedroni1978} on
    $\pi^{+} p$ cross-section in energy region from threshold up to $1.32\;\GeV$.}
\end{figure}

\begin{figure}[h]
  \includegraphics*[width=7cm]{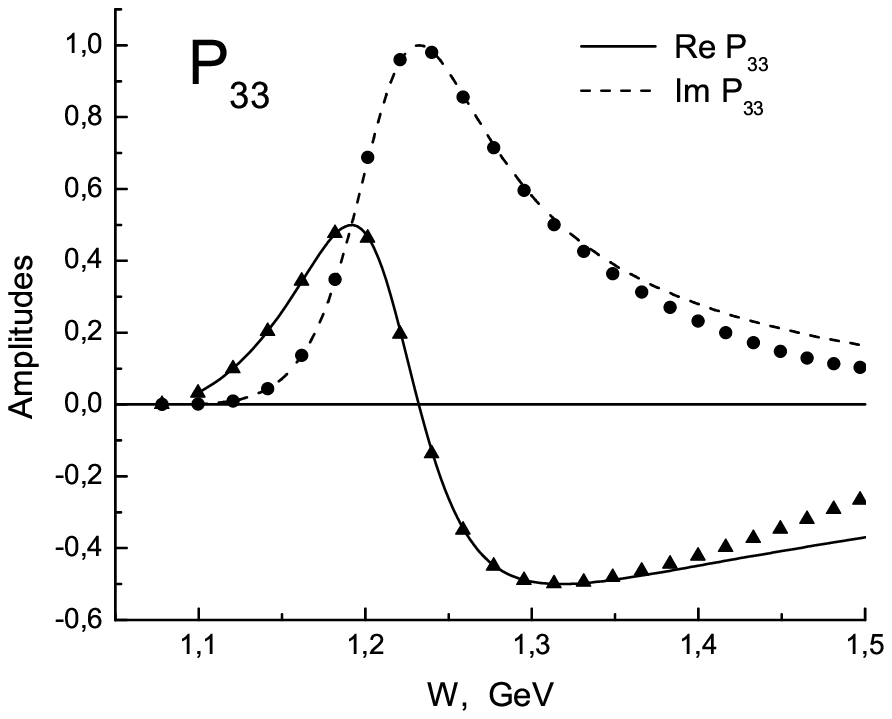}
  \includegraphics*[width=7cm]{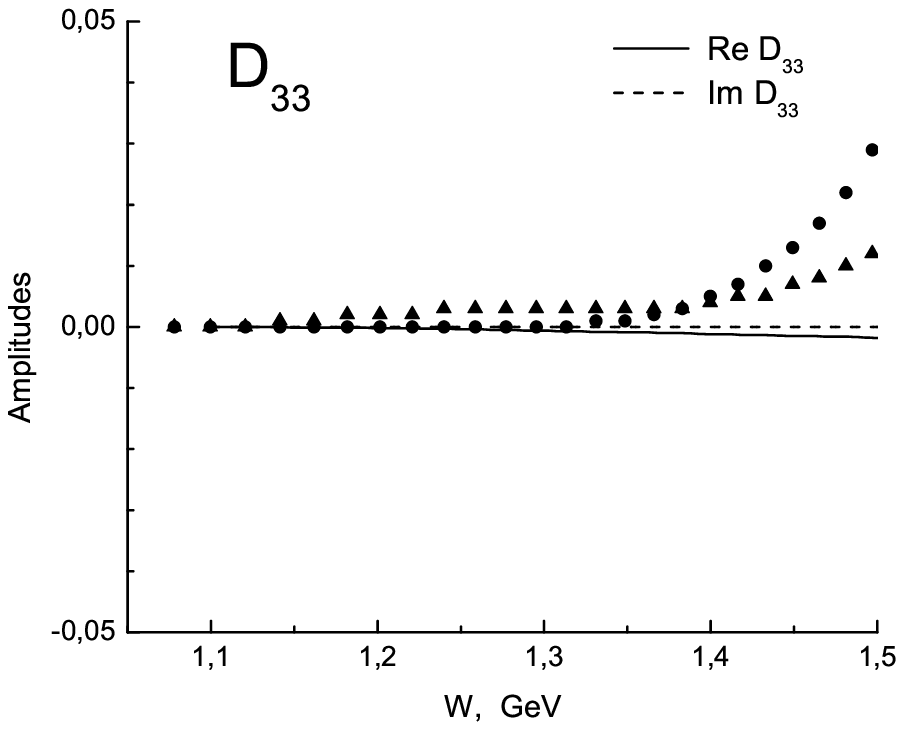}\\
  \includegraphics*[width=7cm]{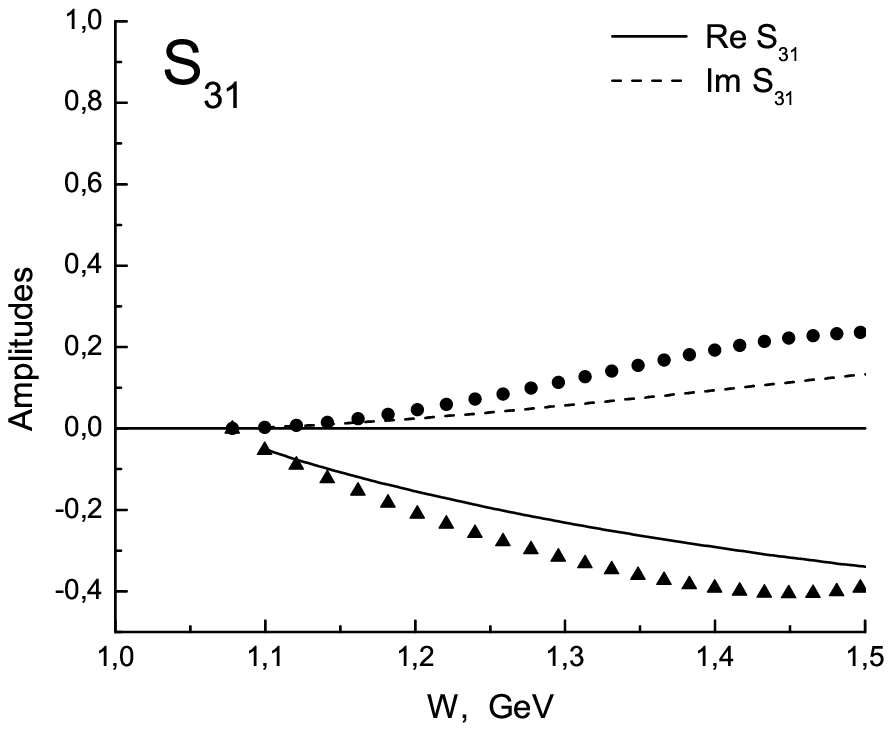}
  \includegraphics*[width=7cm]{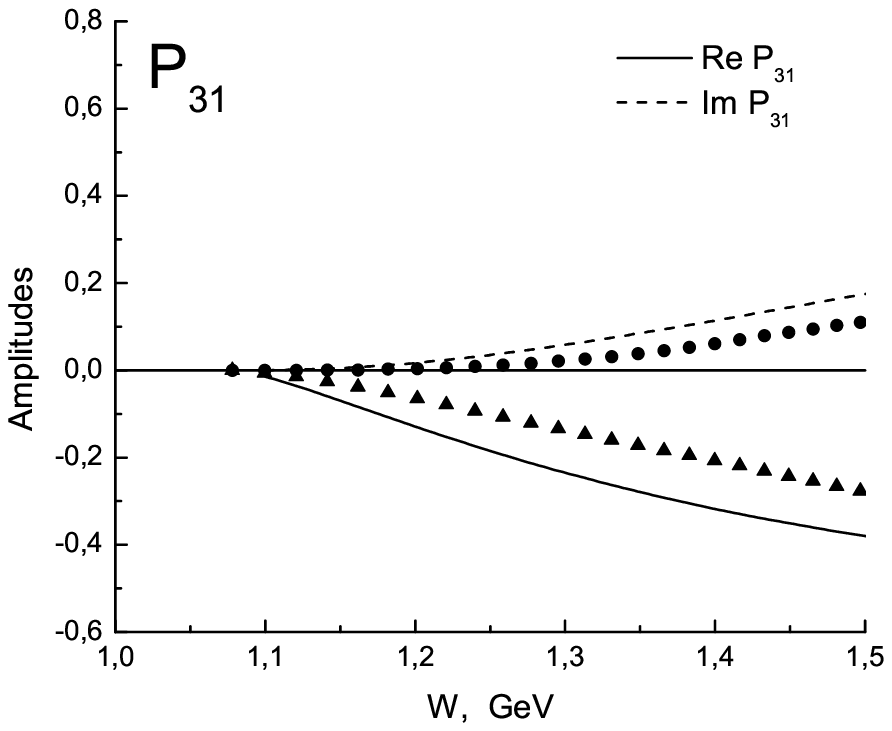}\\
  \caption{\label{fig:PW} Comparison of our partial waves with results of partial wave analysis
    \cite{Arndt2006} (they are very close to results of previous analysis \cite{Hoh1983,Cut1979}
    in this region). Parameters in our amplitudes are taken from total cross section fit
    \eqref{eq:best}. Our partial waves satisfy the elastic unitary condition $\Im
    f=\abs{f}^{2}$.}
\end{figure}

\begin{figure}[h]
  \includegraphics*[width=7cm]{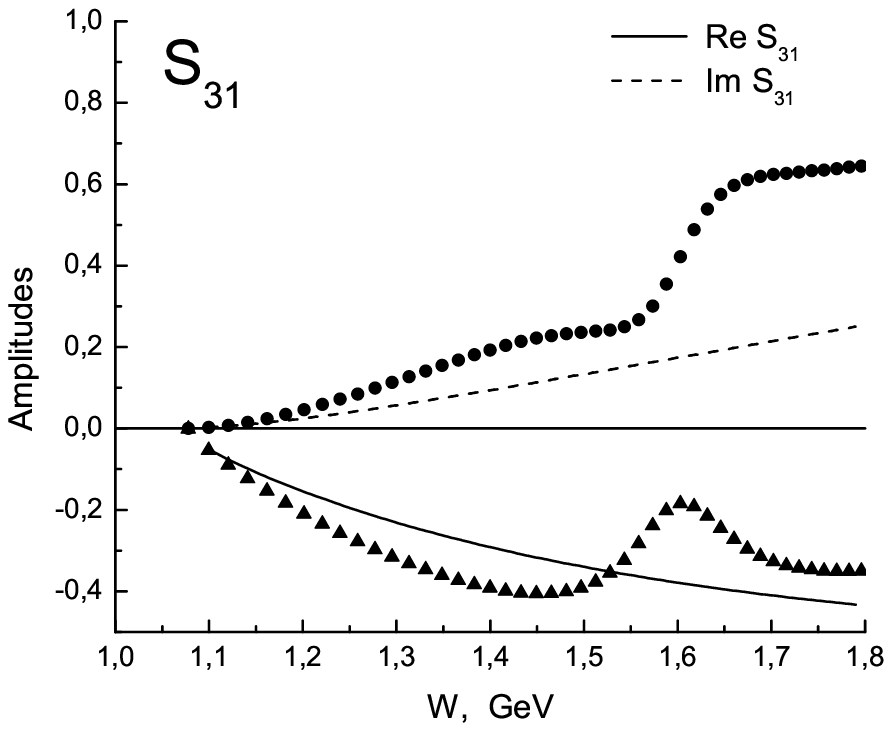}
  \includegraphics*[width=7cm]{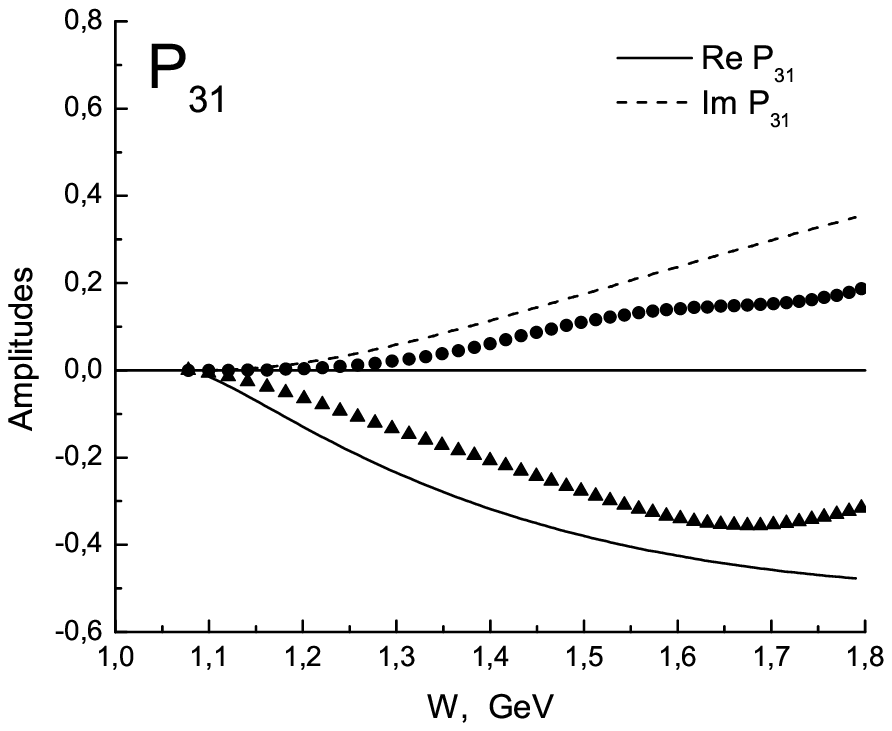}\\
  \caption{\label{fig:PW2} The same as Fig.~\ref{fig:PW} for partial waves $S_{31}$ and $P_{31}$
    in extended energy region.}
\end{figure}


\end{document}